\documentclass{article}

\usepackage{arxiv}

\usepackage[utf8]{inputenc} 
\usepackage[T1]{fontenc}    
\usepackage{hyperref}       
\usepackage{url}            
\usepackage{booktabs}       
\usepackage{amsfonts}       
\usepackage{nicefrac}       
\usepackage{microtype}      
\usepackage{lipsum}
\usepackage{graphicx}
\graphicspath{ {./images/} }

\usepackage{algorithm}
\usepackage{algorithmic}
\usepackage{appendix}
\usepackage[most]{tcolorbox}
\usepackage{xcolor}
\usepackage{booktabs}
\usepackage{amsmath,amsthm,amssymb}
\usepackage{bbold}
\newtheorem{definition}{Definition}
\newtheorem{theorem}{Theorem}

\usepackage{bbding}

\definecolor{promptbackground}{HTML}{F8F8F8} %
\definecolor{promptborder}{HTML}{DDDDDD}
\definecolor{prompttitle}{HTML}{444444}

\usepackage{newfloat}
\usepackage{listings}
\usepackage{threeparttable}

\title{LLaPipe: LLM-Guided Reinforcement Learning for Automated Data Preparation Pipeline Construction}

\newcommand{\szucsse}{$^{1}$}
\newcommand{\szuai}{$^{2}$}
\author{
 Jing Chang \szucsse \\
  \texttt{2350273007@email.szu.edu.cn} \\
   \And
 Chang Liu \szucsse \\
  \texttt{2300271084@email.szu.edu.cn} \\
  \And
 Jinbin Huang \szuai \\
  \texttt{jbhuang@comp.hkbu.edu.hk} \\
  \And
 Rui Mao \szucsse \\
  \texttt{mao@szu.edu.cn} \\
  \And
 Jianbin Qin \szucsse \textsuperscript{\Envelope} \\
  \texttt{qinjianbin@szu.edu.cn} \\
  \And \\
    \szucsse College of Computer Science and Software Engineering, Shenzhen University \\ \szuai School of Artificial Intelligence, Shenzhen University
}

\begin{document}
\maketitle
\begin{abstract}
Automated data preparation is crucial for democratizing machine learning, yet existing reinforcement learning (RL) based approaches suffer from inefficient exploration in the vast space of possible preprocessing pipelines. We present LLaPipe, a novel framework that addresses this exploration bottleneck by integrating Large Language Models (LLMs) as intelligent policy advisors. Unlike traditional methods that rely solely on statistical features and blind trial-and-error, LLaPipe leverages the semantic understanding capabilities of LLMs to provide contextually relevant exploration guidance. Our framework introduces three key innovations: (1) an LLM Policy Advisor that analyzes dataset semantics and pipeline history to suggest promising preprocessing operations, (2) an Experience Distillation mechanism that mines successful patterns from past pipelines and transfers this knowledge to guide future exploration, and (3) an Adaptive Advisor Triggering strategy (Advisor\textsuperscript{+}) that dynamically determines when LLM intervention is most beneficial, balancing exploration effectiveness with computational cost. Through extensive experiments on 18 diverse datasets spanning multiple domains, we demonstrate that LLaPipe achieves up to 22.4\% improvement in pipeline quality and 2.3× faster convergence compared to state-of-the-art RL-based methods, while maintaining computational efficiency through selective LLM usage (averaging only 19.0\% of total exploration steps). The codes and datasets are available at (xxx).
\end{abstract}


\section{Introduction}\label{sec:intro}
 \label{sec:introduction}
In the modern machine learning lifecycle, the process of data preparation constitutes a foundational, yet profoundly challenging phase. This stage, encompassing a diverse set of tasks including data cleansing, transformation, feature engineering, and selection, serves as the bedrock upon which all subsequent modeling efforts are built. The quality of this preparatory work directly and decisively dictates the performance, accuracy, and robustness of the final machine learning models~\cite{pecan2024data}. Despite its paramount importance, data preparation has historically remained a largely manual, intuition-driven, and labor-intensive endeavor. Industry analyses consistently report that data scientists dedicate between 50\% and 80\% of their project time to these tasks~\cite{minh2018automated}, highlighting a significant operational bottleneck that impedes the scalable deployment of machine learning solutions~\cite{John2025data}.

The advent of Automated Machine Learning (AutoML) has sought to address this challenge by automating the end-to-end machine learning workflow, from raw data to a deployable model. Existing methods include rule-based systems, template-based pipelines, and general search heuristics such
as genetic algorithms~\cite{olson2016evaluation,siddiqi2023saga}, Bayesian optimization~\cite{feurer2015efficient,shang2019democratizing,thornton2013auto} and differential formulation~\cite{hilprecht2023diffml,li2023diffprep,yu2021windtunnel}. A particularly promising approach within this domain is the application of Reinforcement Learning (RL)~\cite{berti2019learn2clean,chen2023haipipe,heffetz2020deepline,shang2019democratizing,yang2021auto,gao2024ctxpipe,koka2025cleansurvival,chen2025advancing}. The core innovation of using RL is to reframe the entire data preparation process as a sequential decision-making problem.
This approach models the process as a Markov Decision Process (MDP)~\cite{mdp},
the RL agent's objective is to learn a policy for selecting operators that maximizes cumulative reward~\cite{minh2018automated}. This paradigm has the potential to transform data preparation from a bespoke art form into a rigorously solvable optimization problem, discovering novel and effective data transformation pipelines that may elude human intuition~\cite{DBLP:journals/corr/abs-1709-07150}. Recent advances, such as CtxPipe~\cite{gao2024ctxpipe}, have pushed the state-of-the-art by integrating Deep Q-Networks (DQN) with data context information to guide the search process more effectively.

\begin{figure*}[t]
    \centering
    \includegraphics[width=\linewidth]{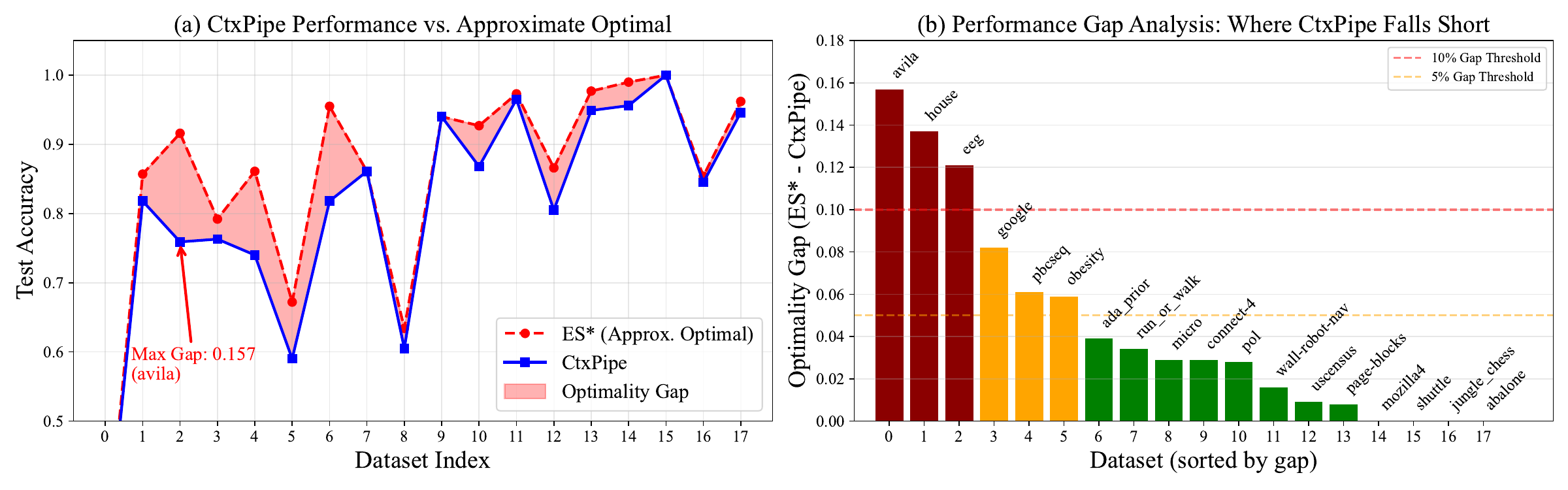}
    \caption{Performance gap analysis showing CtxPipe vs ES* accuracy across 18 datasets}
    \label{fig:Performance gap analysis}
\end{figure*}

However, despite these advances, a deeper analysis reveals that even state-of-the-art methods are constrained by fundamental limitations in their intelligent exploration capabilities. Figure~\ref{fig:Performance gap analysis} presents a revealing analysis of CtxPipe's performance compared to an approximated optimal solution (ES*) obtained through extensive exhaustive search with 10,000 iterations.

The results expose a concerning pattern: despite incorporating contextual information and sophisticated deep learning mechanisms, CtxPipe exhibits substantial optimality gaps across multiple datasets. Most notably~\footnote{All experimental data and analysis presented are directly drawn from CtxPipe~\cite{gao2024ctxpipe}.}:
\begin{itemize}
    \item Systematic underperformance: On datasets like \texttt{avila} and \texttt{eeg}, CtxPipe achieves only 75.9\% and 74.0\% accuracy respectively, while the approximate optimal reaches 91.6\% and 86.1\% – max gap exceeding 15\%.
    \item Computational inefficiency: Even with context-aware guidance, CtxPipe requires extensive training (32,000 steps) yet still misses superior pipeline configurations that exhaustive search eventually discovers.
    \item Inconsistent exploration: The variance in optimality gaps (ranging from 0 to 15.7\%) suggests that the exploration strategy fails to adapt to different data characteristics.
\end{itemize}

\begin{table}[htbp]
    \centering
    \caption{Comparison of performances on \texttt{avila} dataset}
    \begin{tabular}{ccl}
    \toprule
        Approach & Accuracy & Pipeline (with operator id\tnote{1}) \\ \midrule
        CtxPipe & 0.759 & [24, 12, 23] \\
        ES\textsuperscript{*} & 0.916 & [10, 23, 18] \\
        \bottomrule
    \end{tabular}
    \begin{tablenotes}
        \footnotesize
        \item[1] The (id, operator\_name) mappings are listed in Table~\ref{tab:operator_mapping}.
    \end{tablenotes}
    \label{tab:avila-compare}
\end{table}

This performance gap is starkly illustrated by the \texttt{avila} dataset as Table~\ref{tab:avila-compare} listed, CtxPipe consistently selects standard preprocessing pipelines (VarianceThreshold → PowerTransform → RandomTreesEmbedding). However, the superior pipeline discovered by exhaustive search employs a counter-intuitive combination (QuantileTransformer → RandomTreesEmbedding → PCA\_LAPACK) that better captures the data's unique features. 

The agent's inability to explore such unconventional combinations, despite having contextual information about the data domain, underscores two fundamental challenges that limit current automated data preparation methods: (1) \textbf{Myopic exploration:} Existing RL agents, even with sophisticated architectures like DQN, tend to converge prematurely to suboptimal policies, missing the best preprocessing combinations. (2) \textbf{Inefficient search}: The vast combinatorial space of possible pipelines (with just 5 component types and 4 candidates each, the space exceeds 100,000 configurations) demands more intelligent navigation than current $\epsilon$-greedy or softmax exploration strategies provide. 



To address these limitations, we introduce LLaPipe, a novel framework that addresses these limitations through intelligent, uncertainty-aware exploration guided by large language models. Unlike existing methods that rely solely on statistical features or fixed exploration strategies, LLaPipe leverages the semantic understanding and reasoning capabilities of LLMs to provide dynamic, context-aware guidance to the RL agent. Our key contributions include:
\begin{itemize}
    \item We introduce an LLM-based advisor that provides intelligent guidance to the RL agent. By analyzing data features and historical operational context, the LLM suggests high-potential actions, steering the exploration process away from unpromising regions and towards more relevant parts of the search space, thus enhancing search efficiency. 
    \item We design a mechanism to distill reusable knowledge from successful pipeline executions. This component mines the Experience Pool for frequent patterns and high-performing operator sequences, abstracting them into generalizable rules that inform the LLM Advisor, enabling cross-task knowledge transfer.
    \item Recognizing that constant LLM intervention can be costly and sometimes unnecessary, we develop an adaptive triggering mechanism. This component monitors the RL agent's learning progress and dynamically decides when to invoke the LLM Advisor, balancing the cost of LLM guidance with the agent's learning efficiency.
\end{itemize}


\section{Related Works}\label{sec:related}
\subsection{Automated Data Preparation and AutoML}
The challenge of automating data preparation has garnered significant attention in the machine learning community~\cite{yu2018learning}. Early AutoML systems focused primarily on model selection and hyperparameter optimization, with tools like Auto-WEKA~\cite{thornton2013auto} and Auto-sklearn~\cite{feurer2015efficient} pioneering the automated selection of learning algorithms and their configurations. TPOT~\cite{olson2016evaluation} extended this paradigm by using genetic programming to automatically design and optimize complete machine learning pipelines, including preprocessing steps.
More recent work has recognized data preparation as a critical bottleneck. \cite{shang2019democratizing}~developed an interactive system for democratizing data science through ML pipeline curation, while \cite{siddiqi2023saga} introduced SAGA, a scalable framework specifically targeting the optimization of data cleaning pipelines. These systems demonstrate the growing recognition that data preparation quality directly impacts downstream model performance.

\subsection{Reinforcement Learning for Pipeline Construction}
The application of reinforcement learning to automated data preparation represents a paradigm shift in how we approach pipeline construction. \cite{khurana2018feature} were among the first to frame feature engineering as a reinforcement learning problem, demonstrating that RL agents could learn effective transformation sequences~\cite{yu2018towards,zhang2025causalcomrl}. This foundational work inspired several subsequent approaches.
Learn2Clean (Berti-Equille 2019) applied RL to optimize sequences of data cleaning tasks for web data, while Minh et al. (2018) extended RL to automated image preprocessing. More sophisticated approaches emerged with DeepLine~\cite{heffetz2020deepline}, which combined deep reinforcement learning with hierarchical action filtering to manage the vast action space of possible preprocessing operations.
Recent advances include HAIPipe~\cite{chen2023haipipe}, which innovatively combines human-generated and machine-generated pipelines, leveraging both human expertise and automated search. \cite{yang2021auto} introduced Auto-pipeline, using RL with search techniques to synthesize complex data pipelines. The state-of-the-art CtxPipe~\cite{gao2024ctxpipe} significantly advanced the field by incorporating data context embeddings into Deep Q-Networks, enabling more informed preprocessing decisions. 

\subsection{LLM-Guided Optimization in automation tasks}
To address this exploration challenge, we turn to Large Language Models (LLMs), whose reasoning capabilities are transforming various automation tasks. \cite{narayan2022can,xiong2024watch} developed a system where LLMs suggest data transformations based on column semantics, though limited to single-step recommendations. \cite{hegselmann2023tabllm,shoaeinaeini2024guiding} proposed TabLLM for automated feature engineering, using LLMs to generate semantic features from tabular data. Most recently, \cite{borisov2022deep} surveyed the emerging field of LLMs for tabular data, highlighting the potential for semantic understanding in data preprocessing tasks. However, the application of LLMs to guide reinforcement learning agents in automated data preparation remains largely unexplored. Our work pioneers this integration by using LLMs not just for one-shot predictions, but as strategic advisors that provide contextual guidance throughout the RL exploration process.

\section{Preliminaries}\label{sec:prelimnaries}


\begin{definition}[Data Preparation Pipeline]
A data preparation pipeline, $P$, is an ordered sequence of operators, $(a^{(0)}, a^{(1)}, \dots, a^{(L-1)})$, applied to an initial dataset $X_0$. Each operator $a^{(t)}$ is chosen from a predefined set of available operators $\mathcal{O}$. The application of the full pipeline produces a final transformed dataset $X_L$:\normalfont
\begin{equation}
    X_{t+1} = a^{(t)}(X_t) \quad \text{for } t = 0, 1, \dots, L-1
\end{equation}
\end{definition}


\begin{definition}[MDP for Pipeline Construction]
We frame the construction of an optimal pipeline as finding a policy for a Markov Decision Process (MDP). A state $s_t$ is a \textbf{vector representation} summarizing the current dataset and the operators applied so far. An action $a_t$ is the choice of a data preparation operator from a set $\mathcal{O}$ or a special termination action. The agent receives a reward $r_t$, which is primarily based on the downstream performance of the final pipeline. The goal is to learn an optimal policy $\pi^*: \mathcal{S} \to \mathcal{A}$ that maximizes the expected cumulative reward.
\end{definition}
\begin{definition}[Experience Entry]
\label{def:experience_entry}
An experience entry, $e$, stored in the Experience Pool $\mathcal{E}$, is a composite object designed for multi-level, similarity-based retrieval. Each entry contains: \textbf{1) a global summary}, $(D_{\text{meta\_vec}}, P_{\text{complete}}, R_{\text{final}})$, which includes a vector of the initial dataset's features, the final pipeline, and its performance; and \textbf{2) a step-wise trajectory}, $\{(s_{i\_\text{vec}}, a_i, r_i)\}_{i=0}^{L-1}$, which records the sequence of intermediate state vectors, actions, and rewards. This dual structure enables retrieval based on both overall dataset similarity and fine-grained, intermediate state similarity.
\end{definition}

\textbf{Problem Statement.}
Given a raw dataset $X_0$, a set of available data preparation operators $\mathcal{O}$, and a downstream machine learning task $\tau$ evaluated by a performance metric $\text{Eval}(\cdot)$, the objective of automated data preparation pipeline construction is to find an optimal pipeline $P^*$. 

An optimal pipeline $P^*$ is a sequence of operators from $\mathcal{O}$ that maximizes the performance metric when applied to $X_0$:
\begin{equation}
    P^* = \underset{P \in \mathcal{P}_{\text{valid}}}{\arg\max} \left( \text{Eval}(\tau, P(X_0)) \right)
    \label{eq:problem_statement}
\end{equation}
where $P(X_0)$ denotes the dataset resulting from applying pipeline $P$ to $X_0$, and $\mathcal{P}_{\text{valid}}$ is the space of all valid pipelines that can be constructed from $\mathcal{O}$ up to a maximum length $L_{\text{max}}$.

The core challenge lies in efficiently searching the vast and complex combinatorial space $\mathcal{P}_{\text{valid}}$. The size of this space grows exponentially with the number of operators and the pipeline length. Furthermore, the interactions between operators are often non-obvious and highly dependent on the dataset's specific characteristics. A brute-force or naive search is computationally infeasible. Therefore, this problem necessitates an intelligent exploration strategy that can navigate this space efficiently to discover high-quality, and potentially non-intuitive, pipelines.

\section{The LLaPipe Framework}\label{sec:overview}
\subsection{Framework Overview}

\begin{figure*}[t]
    \centering
    \includegraphics[width=0.8\linewidth]{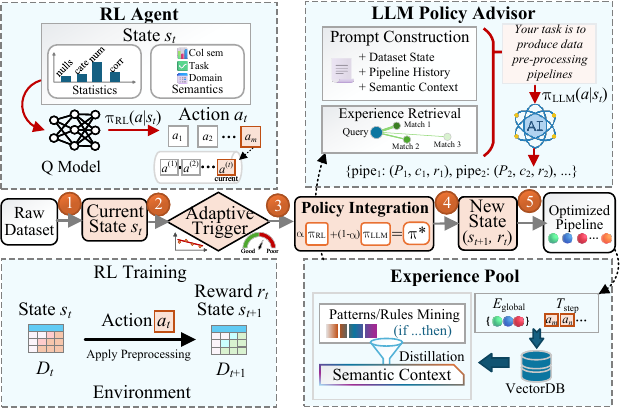}
    \caption{Architecture of LLaPipe, which consists of three key modules (1) RL Agent, (2) LLM Policy Advisor, and (3) Experience Pool, including the general workflow}
    \label{fig:architecture}
\end{figure*}

As illustrated in Figure~\ref{fig:architecture}, LLaPipe augments traditional RL agents with three key innovations: (i) an LLM Policy Advisor that provides semantic-aware exploration guidance, (ii) a sophisticated Experience Management and Distillation mechanism, and (iii) an Adaptive Advisor Triggering strategy that optimizes the cost-benefit tradeoff of LLM intervention.

The general workflow of LLaPipe, from receiving a raw dataset to producing an optimized data preparation pipeline, can be summarized by the following iterative steps, which are elaborated upon in Appendix A.2(Algorithm~\ref{alg:main_llapipe_loop}):
\begin{enumerate}
    \item \textbf{State Construction:} The process begins with a raw dataset. For each step $t$ in the pipeline construction, the system creates a comprehensive \textbf{Current State} representation, $s_t$. 
    \item \textbf{Adaptive Triggering:} The state $s_t$ is passed to the \textbf{Adaptive Trigger}, a crucial decision gate. This module monitors the agent's performance trend. 
    \item \textbf{Policy Generation and Integration:} This stage is the heart of LLaPipe's intelligence.
    \begin{itemize}
        \item If the advisor is not triggered, the RL Agent's \textbf{Q Model} generates its native policy, $\pi_{\text{RL}}(a|s_t)$.
        \item If the advisor is triggered, the \textbf{LLM Policy Advisor} module begins its work. It first performs \textbf{Experience Retrieval}, querying the \textit{Experience Pool} for semantically similar past successes. These retrieved examples, along with the current state and pipeline history, are used in \textbf{Prompt Construction} to create a rich, context-aware prompt for the LLM. The LLM then generates a set of candidate pipelines, forming the advisor's policy, $\pi_{\text{LLM}}(a|s_t)$.
    \end{itemize}
    \item \textbf{Environment Interaction and State Transition:} The action $a_t$ chosen by the integrated policy is executed in the \textbf{Environment}. This involves applying the corresponding preprocessing operator to the dataset $D_t$, resulting in a new dataset $D_{t+1}$ and a reward $r_t$. The system transitions to a \textbf{New State} $(s_{t+1}, r_t)$.
    \item \textbf{Experience Management and Distillation:} The workflow culminates in long-term knowledge curation. Successful trajectories are stored in the \textbf{Experience Pool}, which is implemented as a vector database. 
\end{enumerate}

\subsection{LLM Policy Advisor}
The LLM Policy Advisor acts as the strategic core of LLaPipe,
designed to inject high-level reasoning into the RL agent's exploration process. Its function is to generate contextually-aware, high-quality action recommendations.

\subsubsection{Semantic Prompt Construction}
At each invocation, the advisor constructs a comprehensive prompt for the LLM. This is not a static template but a dynamic composition tailored to the current state $s_t$. The process involves two key steps:
\begin{itemize}
    \item State Encoding: The current state $s_t$, which encapsulates the statistical and semantic features of the dataset $D_t$ and the history of applied operators, is prepared for analysis.
    \item Retrieval-Augmented Generation (RAG)~\cite{gao2023retrieval,lewis2020retrieval}: The vectorized representation of $s_t$ is used as a query to the Experience Pool, a vector database~\cite{pan2024vector} storing historical trajectories. Using efficient vector similarity search, the advisor retrieves the k most relevant past experiences. These ``in-context examples'' consist of historical states that were similar to the current one, along with the successful actions and resulting rewards from those situations, further promot details can be found in Appendix A.4\ref{app:prompt_design}.
\end{itemize}

\subsubsection{LLM-Guided Action Generation}
The LLM processes the structured, retrieval-augmented prompt to reason about the optimal next steps. Its output is not a single action but a ranked list of candidate actions or short action sequences, $\mathcal{A}_{\text{suggested}} = \{(P_i, c_i, r_i)\}_{i=1}^k$, where $P_i$ is a suggested preprocessing pipeline, $c_i \in [0, 1]$ is the LLM's confidence in that suggestion, and $r_i$ is the textual rationale.
\subsubsection{Policy Integration Mechanism}
To balance the LLM's strategic guidance with RL agent's learned experience, we formulate a hybrid policy, $\pi_{\text{combined}}$, the final action $a_t$ is chosen based on a weighted combination of the LLM's suggestions and the agent's native policy:
\begin{equation}
\pi_{\text{combined}}(a|s_t) = \alpha \cdot \pi_{\text{LLM}}(a|s_t) + (1 - \alpha) \cdot \pi_{\text{RL}}(a|s_t)
\label{eq:policy_integration}
\end{equation}
where:
\begin{itemize}
    \item $\pi_{\text{LLM}}(a|s_t)$ is derived from the LLM's confidence scores, typically as a normalized distribution: $\pi_{\text{LLM}}(a|s_t) \propto c_a$ for $a \in \mathcal{A}_{\text{suggested}}$.
    \item $\pi_{\text{RL}}(a|s_t)$ is the agent's native policy, e.g., a softmax distribution over its learned Q-values: $\pi_{\text{RL}}(a|s_t) \propto \exp(Q(s_t, a)/\tau)$.
    \item $\alpha \in [0, 1]$ is a dynamic weighting factor determined by the Adaptive Advisor Triggering, if the advisor is not triggered, $\alpha$ = 0, else $\alpha$ can be set to a fixed value.
\end{itemize}

\subsection{Experience Distillation}
\label{sec:exp_distillation}
The Experience Distillation module enables LLaPipe to learn from its history, creating a powerful feedback loop that continually enhances the LLM Advisor's guidance.

\subsubsection{Experience Pool Management}
We maintain an Experience Pool, $\mathcal{E}$, implemented as a vector database\cite{han2023comprehensive,jing2025large}.  When a pipeline completes successfully, its trajectory is processed and stored as a new experience entry $e \in \mathcal{E}$. Each entry contains both the global summary and the step-wise details, as detailed in Definition~\ref{def:experience_entry}.
\begin{itemize}
    \item Global Summary ($E\_global$): A tuple ($D\_meta\_vec$, $P\_complete$, $R\_final$) where $D\_meta\_vec$ is the vectorized representation of the initial dataset's features. This allows for retrieving entire pipelines that worked well on similar types of datasets.
    \item Step-wise Trajectory ($T\_stepwise$): A sequence {($s\_i\_vec$, $a\_i$, $r\_i$)} where $s\_i\_vec$ is the vectorized intermediate state. This allows for retrieving successful next steps from similar intermediate situations.
\end{itemize}

\subsubsection{Knowledge Mining and Abstraction}

Periodically, an offline process analyzes the entire Experience Pool to distill higher-level, generalizable knowledge. This goes beyond simple retrieval and involves:

\begin{enumerate}
    \item \textbf{Sequential Pattern Mining:} Using algorithms like PrefixSpan to identify frequently co-occurring sequences of operators (e.g., StandardScaler is often followed by PCA) that consistently lead to high rewards.
    
    \item \textbf{Contextual Rule Mining:} Learning association rules of the form IF (\texttt{dataset\_properties}) THEN (\texttt{apply\_operator\_sequence}), which connect specific data characteristics to optimal strategies.

\end{enumerate}
This distilled knowledge is stored in a separate knowledge base that the LLM Policy Advisor can reference during prompt construction, further enriching its reasoning capabilities.

\subsection{Adaptive Advisor Triggering (Advisor\textsuperscript{+})}
\label{sec:adaptive_triggering}

A primary challenge in integrating LLMs into an RL loop is managing the computational cost and latency of LLM invocations. Naively calling the Advisor at every step is prohibitively expensive. To address this, we introduce the \textbf{Adaptive Advisor Triggering} mechanism, named Advisor\textsuperscript{+}, a sophisticated strategy designed to invoke the LLM Policy Advisor only when its guidance is most likely to be impactful.

\subsubsection{Performance Trend Monitoring}

We maintain a sliding window buffer $\mathcal{B} = \{acc_1, acc_2, \ldots, acc_{|B|}\}$ containing the accuracy scores from the most recent episodes. After each episode $e$, we evaluate the final pipeline's accuracy on the validation set and update:
\begin{equation}
\mathcal{B} \leftarrow \mathcal{B} \cup \{acc_e\} \setminus \{acc_{oldest}\}
\end{equation}
where $|\mathcal{B}| \leq 10$ to focus on recent performance trends.
\subsubsection{Trend Analysis via Linear Regression}

To quantify the performance improvement trajectory, we compute the slope of a linear regression model fitted to the recent accuracies:
\begin{equation}
\beta = \frac{\sum_{i=1}^{|\mathcal{B}|} (i - \bar{i})(acc_i - \overline{acc})}{\sum_{i=1}^{|\mathcal{B}|} (i - \bar{i})^2}
\end{equation}
where $\bar{i}$ and $\overline{acc}$ are the mean indices and accuracies respectively. The slope $\beta$ represents the rate of performance improvement per episode.

The decision to trigger the advisor is not heuristic but is grounded in a theoretical cost-benefit analysis. Our strategy relies on the following theoretical justifications.

First, we justify that a linear slope is a valid local approximation of the learning curve.
\begin{theorem}[Local Linear Approximation]
\label{thm:linear_approx}
The agent's expected accuracy improvement over a short series of episodes can be locally approximated by a linear function, $\mathbb{E}[\text{acc}(e)] \approx \text{acc}_0 + \beta \cdot e$. 
\end{theorem}

This allows us to formulate an optimal triggering policy based on a cost-benefit analysis.
\begin{theorem}[Optimality of Slope-Based Triggering]
\label{thm:slope_optimality}
For a given LLM invocation cost and expected accuracy gain, there exists a critical slope threshold, $\theta_{slope}$, below which invoking the LLM is the optimal action that maximize long-term reward.
\end{theorem}

Furthermore, we optimize the evaluation of the LLM's suggestions to minimize computational cost.
\begin{theorem}[Optimality of First-Improvement Sampling]
\label{thm:first_improvement}
When evaluating a list of LLM-generated pipelines ordered by confidence, the strategy of stopping at the first pipeline that outperforms the baseline is optimal in terms of balancing expected gain and evaluation cost.
\end{theorem}

Finally, the overall efficiency of our adaptive approach is formally established.
\begin{theorem}[Expected Cost Reduction]
\label{thm:cost_reduction}
Compared to a fixed-frequency intervention strategy, Advisor\textsuperscript{+} significantly reduces the total expected computational cost over an entire training run.
\end{theorem}

Our final policy, detailed in Algorithm~\ref{alg:advisor_trigger} in the Appendix A.2, is thus to invoke the advisor if and only if the performance slope $\beta$ falls below the threshold $\theta_{slope}$. All proofs for Theorems~\ref{thm:linear_approx}-\ref{thm:cost_reduction} are provided in Appendix A.6. This adaptive strategy ensures that expert guidance is supplied precisely when needed to escape learning plateaus, ensuring both high final performance and computational efficiency.


\section{Experiments}\label{sec:exp}

\subsection{Experimental Setup}

\noindent\textbf{Datasets.} We evaluate all methods on 18 diverse, real-world datasets from the OpenML repository~\cite{vanschoren2014openml} used by DiffPrep~\cite{li2023diffprep}. These datasets are commonly used benchmarks in AutoML research and span a variety of domains, sizes, and complexities~\cite{li2021cleanml,ledell2020h2o}.

\noindent\textbf{Baselines.} We compare LLaPipe with a wide range of state-of-the-art and standard baseline methods:
\begin{itemize}
    \item \textbf{RL-based Methods:} CtxPipe (the current state-of-the-art method), HAI-AI (applies the deep reinforcement learning-based AI pipeline generation algorithm proposed by HAIPipe~\cite{chen2023haipipe}), Pure DQN algorithm, and Pure Q-Learning algorithm.
    \item \textbf{AutoML/Search-based Methods:} TPOT~\cite{olson2016evaluation}, DeepLine~\cite{heffetz2020deepline}, SAGA~\cite{siddiqi2023saga}.
    \item \textbf{Differentiable Methods:} DP-Fix, DP-Flex from DiffPrep~\cite{li2023diffprep}.
    \item \textbf{LLM-based Methods:} Using LLama3.3-70B~\cite{zheng2024llamafactory} and Qwen3-32B~\cite{yang2025qwen3} for direct pipeline generation.
\end{itemize}

\noindent\textbf{Components \& Case study.} 
Our work utilizes the same set of candidate operators as CtxPipe. Table~\ref{tab:components} in Appendix enumerates the candidate components and their respective types that are utilized in this work. This ensures a fair comparison of the search strategies themselves. However, our framework treats the pipeline construction process with significantly greater flexibility.

A key difference lies in how we handle component selection. While CtxPipe states that the order of component types is not fixed, its implementation imposes constraints that prevent multiple operators of the same type (e.g., two different \textit{Feature Engineering} operators) from co-existing in a single pipeline. This implicitly prunes the search space, potentially missing superior pipeline configurations that rely on such combinations.

In contrast, LLaPipe imposes no such structural priors. The RL agent, guided by the LLM, is free to select any valid operator at any step, allowing for a truly flexible and unconstrained exploration of the pipeline space. This flexibility proves to be a distinct advantage. For instance, on the \texttt{wall-robot-nav} dataset, CtxPipe's constrained search finds a pipeline with an accuracy of 0.946. Our unconstrained approach, however, discovers a superior pipeline: [QuantileTransformer, StandardScaler, PolynomialFeatures], achieving a higher accuracy of 0.961. Notably, this superior pipeline includes two operators, QuantileTransformer and StandardScaler, both belonging to the same \textit{Feature Preprocessing} type—a combination that is inaccessible to CtxPipe's search strategy. This demonstrates LLaPipe's enhanced ability to uncover novel and more effective data preparation sequences.

\noindent\textbf{Model.} Our work utilizes the same downstream ML model Logistic Regression to train and evaluate as CtxPipe~\cite{gao2024ctxpipe} in case of a fair comparison.

\noindent\textbf{Evaluation Metrics.} For the primary task, we report the prediction accuracy of a Logistic Regression model trained on the datasets transformed by the generated pipelines. We also report the average accuracy \textbf{ranking} of each method across all datasets. To evaluate efficiency, we measure \textbf{running time}, \textbf{pipeline length}, and the \textbf{number of LLM calls}.

\subsection{Main Results and Performance Comparison}
\noindent\textbf{Test Accuracy.} Table~\ref{tab:main_results} presents the primary experimental results, we can observe that \textbf{LLaPipe (Advisor)} achieves the best accuracy on 7 out of 18 datasets, achieves the best average accuracy (0.833) and the best average rank (2.67), demonstrating the strong positive impact of LLM guidance. Moreover, by integrating contextual information, under the same component search space, 12 out of 18 datasets get improved accuracy compared with CtxPipe, the largest increased by  22.40\%. Meanwhile, the ranking proportion results are plotted in Figure~\ref{fig:ranks}. \textbf{LLaPipe (Advisor\textsuperscript{+})}, our full model, achieves a very competitive average rank of 2.81, outperforming the state-of-the-art CtxPipe (rank 3.61) and all other baselines, indicating its ability to find high-quality pipelines efficiently.

To provide a more direct and extensive comparison against the current state-of-the-art, we integrated our results into the comprehensive evaluation table presented in the CtxPipe. By placing our LLaPipe variants into this competitive landscape, we can directly assess their performance against a wider array of advanced techniques. The full, extended comparison table is provided in Appendix~\ref{app:full_comparison} (Table~\ref{tab:extended_comparison}). The results further solidify our findings, showing that both LLaPipe variants consistently rank among the top performers.

\begin{figure}
    \centering
    \includegraphics[width=0.5\linewidth]{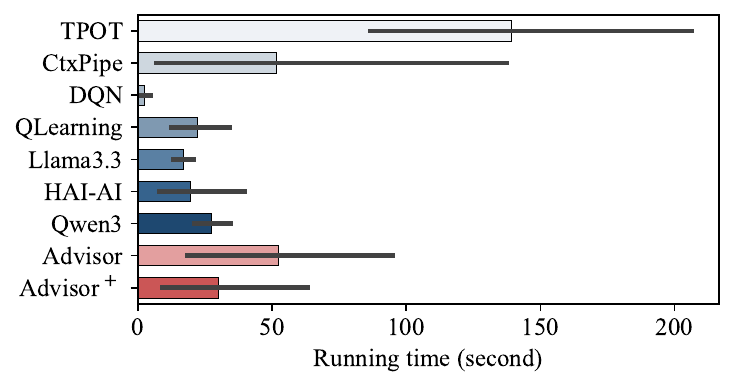}
    \caption{Comparison of running time. The bars indicate the median of running time, and the line on the bar reveals the range of time across the test set.} 
    \label{fig:running_time}
\end{figure}

\noindent\textbf{Running Time and Pipeline Length.} Figure~\ref{fig:running_time} shows that LLaPipe's running time is competitive with other state-of-the-art methods like CtxPipe. Furthermore, 
our methods tend to find shorter, more efficient pipelines (average length 1.89-2.83) compared to the baseline CtxPipe (fixed length of 6 with possible blank operations), further details can be found in Appendix~\ref{app:pipeline_details}(Table~\ref{tab:pipeline_composition}). This suggests that the LLM's guidance helps in finding more parsimonious and effective operator sequences.

\begin{figure}
    \centering
    \includegraphics[width=0.5\linewidth]{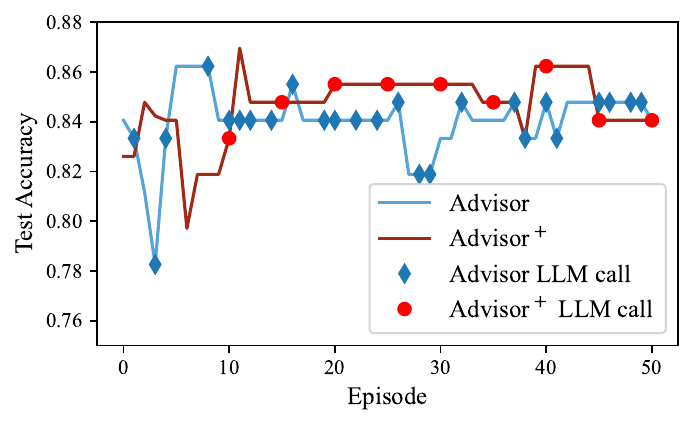}
    \caption{Comparison of LLM invocation times bewteen Advisor and Advisor\textsuperscript{+}}
    \label{fig:llm_calls}
\end{figure}

\begin{table*}[h!]
\centering
\caption{Main experimental results comparing the accuracy of various methods. Advisor and Advisor\textsuperscript{+} are our proposed methods. The best result per dataset is in \textbf{bold}. The rank is averaged over all datasets.}
\label{tab:main_results}
\resizebox{\textwidth}{!}{%
\begin{tabular}{lcccccccccc}
\toprule
\textbf{Dataset} & \textbf{TPOT} & \textbf{CtxPipe} & \textbf{DQN} & \textbf{Q-learning} & \textbf{llama3.3} & \textbf{HAI-AI} & \textbf{Qwen3} & \textbf{Advisor} & \textbf{Advisor\textsuperscript{+}} \\
\midrule
abalone         & \textbf{0.289}& 0.287         & 0.257 & 0.274         & 0.246 & 0.260         & 0.258         & 0.270         & 0.273 \\
ada\_prior      & 0.823         & 0.818         & 0.812 & 0.831         & 0.794 & 0.801         & 0.828         & 0.838         & \textbf{0.841} \\
avila           & 0.593         & 0.759         & 0.620 & 0.682         & 0.613 & 0.630         & 0.542         & \textbf{0.929}& 0.751 \\
connect-4       & 0.658         & 0.763         & 0.664 & 0.747         & 0.657 & \textbf{0.775}& 0.665         & \textbf{0.775}& \textbf{0.775} \\
eeg             & 0.594         & 0.740         & 0.613 & 0.632         & 0.631 & 0.556         & 0.556         & \textbf{0.839}& \textbf{0.811} \\
google          & 0.602         & 0.590         & 0.594 & 0.633         & 0.582 & 0.550         & 0.645         & \textbf{0.675}& \textbf{0.674} \\
house           & \textbf{0.941}& 0.818         & 0.908 & 0.938         & 0.918 & 0.928         & 0.914         & 0.908         & 0.908 \\
jungle\_chess   & 0.682         & \textbf{0.861}& 0.697 & 0.814         & 0.677 & 0.760         & 0.677         & \textbf{0.861}& \textbf{0.861} \\
micro           & 0.558         & 0.605         & 0.589 & 0.613         & 0.609 & 0.633         & 0.578         & \textbf{0.643}& 0.633 \\
mozilla4        & 0.890         & 0.940         & 0.871 & 0.933         & 0.915 & 0.870         & 0.921         & 0.937         & \textbf{0.944} \\
obesity         & \textbf{0.942}& 0.868         & 0.825 & 0.790         & 0.759 & 0.768         & 0.681         & 0.865         & 0.835 \\
page-blocks     & 0.967         & 0.965         & 0.952 & \textbf{0.973}& 0.949 & 0.935         & \textbf{0.973}& 0.965         & 0.961 \\
pbcseq          & 0.715         & \textbf{0.805}& 0.735 & 0.710         & 0.730 & 0.733         & 0.692         & 0.753         & 0.753 \\
pol             & 0.894         & 0.949         & 0.796 & \textbf{0.981}& 0.852 & 0.916         & 0.811         & \textbf{0.981}& \textbf{0.981} \\
run\_or\_walk   & 0.826         & 0.956         & 0.972 & 0.945         & 0.837 & 0.915         & 0.676         & 0.953         & \textbf{0.976} \\
shuttle         & 0.933         & \textbf{1.000}& 0.965 & 0.995         & 0.972 & 0.951         & 0.976         & 0.987         & 0.986 \\
uscensus        & 0.828         & 0.845         & 0.832 & 0.826         & 0.818 & 0.807         & \textbf{0.848}& 0.846         & 0.847 \\
wall-robot-nav & 0.754          & 0.946         & 0.853 & 0.924         & 0.843 & 0.896         & 0.815         & \textbf{0.961}& 0.946 \\
\midrule
\textbf{AVERAGE} & 0.749 & 0.806 & 0.753 & 0.791 & 0.745 & 0.760 & 0.725 & \textbf{0.833} & 0.820 \\
\textbf{RANK} & 5.94 & 3.61 & 6.28 & 4.03 & 7.08 & 6.22 & 6.36 & \textbf{2.67} & \textbf{2.81} \\
\textbf{DATASETS WON} & 3 & 3 & 0 & 2 & 0 & 1 & 2 & \textbf{8} & \textbf{6} \\
\bottomrule
\end{tabular}
}
\end{table*}

\begin{figure}
    \centering
    \includegraphics[width=0.5\linewidth]{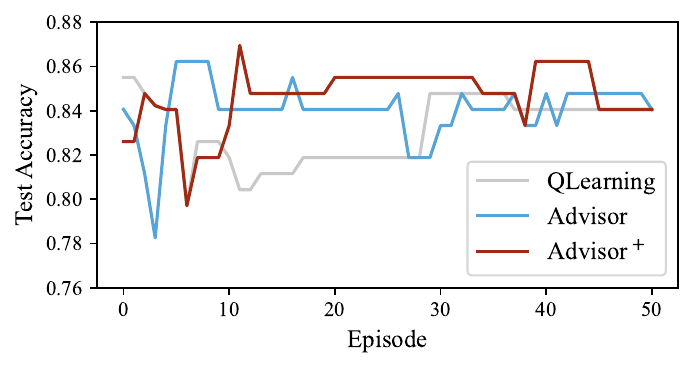}
    \caption{Model performance improves in early 5 to 15 episodes with Advisor, but increases slower in 30 episodes with Q-Learning}
    \label{fig:effectiveness}
\end{figure}

\noindent\textbf{Effectiveness of LLM Guidance.} We first compare the performance of a standard RL (Q-Learning) agent against our LLaPipe (Advisor) variant. As illustrated in Figure~\ref{fig:effectiveness}, the introduction of LLM guidance significantly boosts both the learning speed and the final convergence performance. The LLaPipe (Advisor) curve consistently stays above the baseline RL curve, confirming that the contextual advice provided by the LLM effectively steers the agent out of local optima and towards better solutions.

\noindent\textbf{Evaluation and Analysis of Efficiency (Advisor\textsuperscript{+}).} Once the LLM provides a list of candidate pipelines, we evaluate them sequentially. To optimize computational resources, we employ a ``first-improvement'' sampling strategy, as justified by Theorem~\ref{thm:first_improvement}. We evaluate the pipelines in order of the LLM's confidence and stop as soon as we find one that outperforms the agent's current best performance. The successful trajectory is then used to update the RL agent's knowledge. This evaluation and integration process is formally described in Algorithm~\ref{alg:pipeline_evaluation} in the Appendix A.2.

We further analyze the trade-offs involved in our framework regarding efficiency and the nature of the generated pipelines. A key claim of our work is that the adaptive triggering mechanism (Advisor\textsuperscript{+}) improves efficiency without significantly sacrificing performance. We compare the non-adaptive Advisor with the adaptive Advisor\textsuperscript{+}. Figure~\ref{fig:llm_calls} shows the cumulative number of LLM calls over 50 episodes. The Advisor\textsuperscript{+} variant makes far fewer calls (9 calls) compared to the Advisor (25 calls). This confirms that Advisor\textsuperscript{+} successfully avoids unnecessary LLM invocations when the agent is learning effectively on its own, leading to a significant reduction in computational cost and latency.

\noindent\textbf{Closing the Optimality Gap}
\label{sssec:closing_gap} In the Introduction, we highlighted a significant limitation of existing state-of-the-art methods through Figure~\ref{fig:Performance gap analysis}, which revealed a substantial optimality gap between CtxPipe's performance and an approximated optimal solution (ES*). The analysis showed that CtxPipe, despite its context-awareness, often converges to suboptimal pipelines.

To demonstrate LLaPipe's effectiveness in closing this gap, we revisit this analysis by incorporating our \textbf{LLaPipe (Advisor and Advisor\textsuperscript{+})} into the comparison. Figure~\ref{fig:ctx_llapipe_optim} in Appendix presents a direct comparison of the optimality gap for the most challenging datasets. The results are compelling. On the very datasets where CtxPipe struggled most, LLaPipe significantly narrows or even eliminates the performance gap. This success is rooted in LLaPipe's fundamentally different exploration strategy.


\subsection{Ablation Studies}

To dissect the contribution of each component, we conduct a series of ablation studies to test the efficacy of involving LLM guidance and experience. We first remove the Experience Distillation (ED) module from both Advisor and Advisor\textsuperscript{+}, and secondly, we completely remove the LLM Policy Advisor, which simply applies Q-Learning as the data preparation pipeline constructor.

\textbf{Experience Distillation.} To ablate the ED module, we only employ LLM with semantic data states, available operators, and the pipeline in the last episode, denoted as Advisor w/o ED (or Advisor\textsuperscript{+} w/o ED). The results are presented in Table~\ref{tab:ablation}, key metrics include test accuracy of downstream model ($\uparrow$) and its rank ($\downarrow$), which indicates that with histories and experiences retrieved from the ED module, the LLM could better summarize the context and produce better potential pipelines. In particular, the average accuracy of the downstream task obtained some increase, and the average ranks decrease from 2.81 to 2.67 for Advisor, 3.03 to 2.81 for Advisor\textsuperscript{+}, respectively.

\textbf{LLM Policy Advisor. } We conduct an experiment that only employs Q-Learning to generate pipelines. Without LLM suggestions, the RL agent often fails to explore and find the pipelines in limited episodes, and the performance significantly decreases to 0.791. This reveals that LLM advisor would instruct the RL agent to explore potential trajectories and optimize the model more efficiently.

\begin{table}[htbp]
\centering
\caption{Ablation study}\label{tab:ablation}
\begin{tabular}{lcc}
\toprule
Settings                            & Avg Accuracy   & Avg Rank      \\ \midrule
Advisor                             & \textbf{0.833} & \textbf{2.67} \\
Advisor\textsuperscript{+}          & \textbf{0.820} & \textbf{2.81} \\ \midrule
Advisor (w/o ED)                    & 0.810          & 2.81          \\
Advisor\textsuperscript{+} (w/o ED) & 0.802          & 3.03          \\ \midrule
QLearning                           & 0.791          & 4.03          \\ \bottomrule
\end{tabular}
\end{table}

\section{Conclusion}\label{sec:conclusion}

In this paper, we presented LLaPipe, a novel framework that enhances automated data preparation by integrating Large Language Models (LLMs) as strategic advisors for a Reinforcement Learning agent. By leveraging LLM-driven, context-aware guidance, LLaPipe effectively navigates the vast search space of pipelines. Our key innovations include a retrieval-augmented LLM Policy Advisor, a theoretically-grounded Adaptive Advisor Triggering~(Advisor\textsuperscript{+}) mechanism for efficiency, and a truly flexible search strategy. Extensive experiments show that LLaPipe consistently outperforms state-of-the-art baselines. It achieves superior accuracy while drastically reducing expensive LLM calls, demonstrating its ability to discover novel and more effective pipeline structures.

Our work confirms that guiding RL with LLM reasoning is a powerful paradigm for complex AutoML tasks. Future work will focus on extending LLaPipe to support non-linear pipeline structures and developing self-evolving mechanisms to continuously refine the LLM's advisory capabilities.

\bibliographystyle{unsrt}  
\bibliography{main}  






\appendix
\section{Appendix}
\subsection{Implementation Details}
\label{app:implementation}
\textbf{Hardware and OS.} We run our experiments on a server with 128 AMD EPYC 7543 CPUs, each with 32 cores, and 256GB memory in total. The local LLMs run on NVIDIA A100 with CUDA 12.8. The OS is Ubuntu 20.04 with Linux kernel 5.15.0-138-generic. The Python version of our experiment is 3.12.9.

\noindent\textbf{LLaPipe Variants.} To demonstrate the effectiveness of our contributions, we evaluate two main variants of our framework:
\begin{itemize}
    \item \textbf{LLaPipe (Advisor):} A version that uses LLM guidance at a fixed, frequent interval, without the adaptive triggering mechanism but with experience distillation. This helps isolate the benefit of LLM guidance itself.
    \item \textbf{LLaPipe (Advisor\textsuperscript{+}):} The full proposed framework, including the adaptive triggering (Advisor\textsuperscript{+}) and experience distillation.
\end{itemize}

\noindent\textbf{Q-Learning Agent and State Representation.}
Our RL agent is built upon a traditional, tabular Q-Learning algorithm~\cite{watkins1992q}. A key design choice, is our formulation of the state space. We define the state as the last executed data preparation operator. The action space consists of all available operators that can be applied next.

This approach transforms the complex problem of pipeline construction into navigating a directed graph where nodes represent operators (states) and edges represent the Q-values of transitioning from one operator to another. The Q-table is thus a matrix of size $|\mathcal{O}| \times |\mathcal{O}|$, where $\mathcal{O}$ is the set of all available operators. The Q-value, $Q(s, a)$, represents the expected future reward of applying operator $a$ after having just applied operator $s$.

The Q-table is updated using the classic Bellman equation:
\begin{equation}
    Q(s, a) \leftarrow (1-\alpha)Q(s, a) + \alpha \left( r + \gamma \max_{a'} Q(s', a') \right)
\end{equation}
where $s'$ is the new state (i.e., the action $a$ just taken), $r$ is the reward obtained, $\alpha$ is the learning rate, and $\gamma$ is the discount factor.

\noindent\textbf{Why Q-learning?}
Our choice of a traditional Q-Learning algorithm over more complex Deep Q-Networks (DQN) is a deliberate design decision aimed at ensuring experimental clarity and robustness. This approach offers two key advantages for our study
\begin{itemize}
    \item Isolating the LLM's Impact: A simpler RL foundation allows us to unambiguously attribute performance gains to our core contribution—the LLM Policy Advisor. This avoids confounding variables from the representational power of a deep neural network, allowing for a clear validation of our central hypothesis.
    \item Stability and Reduced Overhead: Unlike DQN, which is notoriously sensitive to hyperparameters and prone to instability in non-stationary environments, Q-Learning provides a more stable and predictable learning process. This significantly reduces engineering overhead and enhances the reproducibility of our results.
\end{itemize}

\noindent\textbf{LLM Policy Advisor and Experience Pool.}
Our LLM Policy Advisor utilizes the \texttt{LLaMA3.3-70B} model as its reasoning engine. The Experience Pool, which stores and retrieves successful trajectories, is implemented using \textbf{FAISS}, a high-performance vector similarity search library. For retrieval, we use the `IndexFlatL2' index, which performs exact, exhaustive search to find the most relevant past examples for the RAG-based prompt construction.

\noindent\textbf{Hyperparameters.}
The specific hyperparameters for each module, including the Q-Learning agent, the LLM, and the Advisor\textsuperscript{+} mechanism, are detailed in Appendix, table~\ref{tab:hyperparameters}. All settings were chosen to ensure a robust and fair comparison across all experiments.

\begin{figure}
    \centering
    \includegraphics[width=0.6\linewidth]{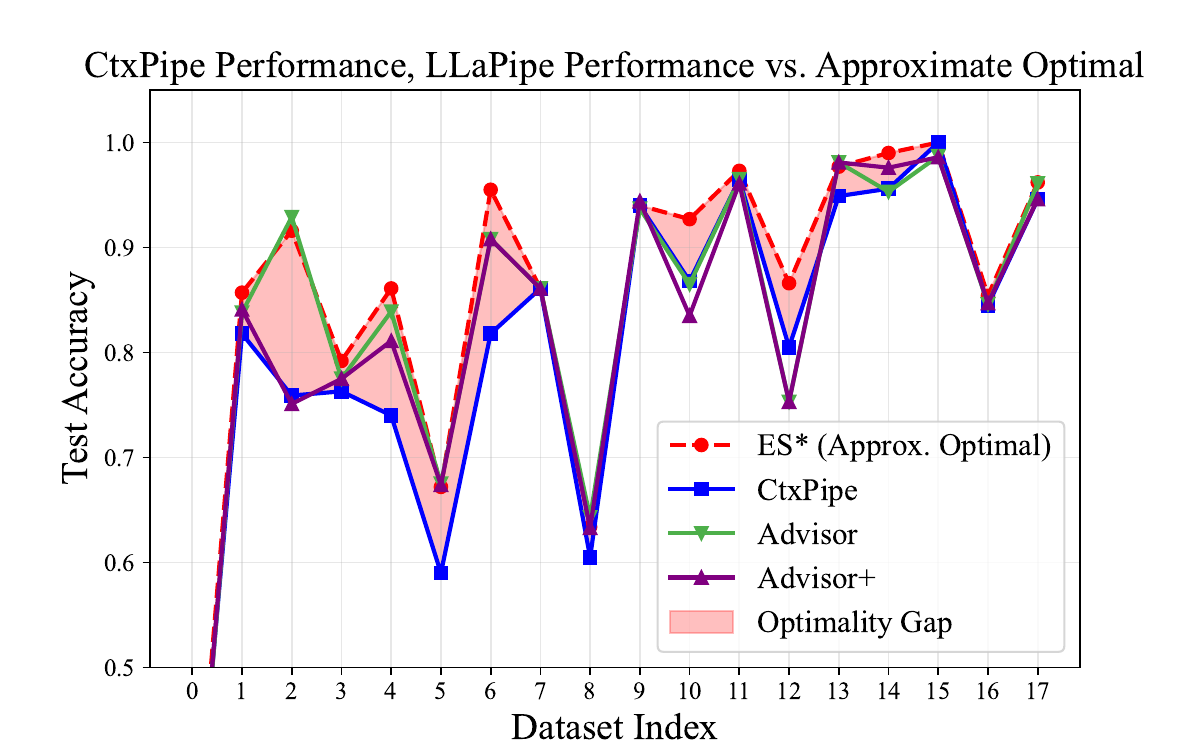}
    \caption{The direct comparison result of the optimality gap
in CtxPipe, LLaPipe and Approximate Optimal}
    \label{fig:ctx_llapipe_optim}
\end{figure}

\begin{table*}[h!]
\centering
\caption{Selected data preparation components and their types}
\label{tab:components}
\begin{tabular}{ll}
\toprule
\textbf{Type} & \textbf{Component} \\
\midrule
\textbf{Imputer} & Mean, Median, Most Frequent \\
\addlinespace %
\textbf{Encoder} & Label Encoder, One-hot Encoder \\
\addlinespace %
\textbf{Feature Preprocessing} & Min Max Scaler, Max Absolute Scaler, Robust Scaler, \\
& Standard Scaler, Quantile Transformer, Log Transformer, \\
& Power Transformer, Normalizer, K-bins Discretizer \\
\addlinespace %
\textbf{Feature Engineering} & Polynomial Features, Interaction Features, PCA, \\
& Kernel PCA, Incremental PCA, Truncated SVD, \\
& Random Trees Embedding \\
\addlinespace %
\textbf{Feature Selection} & Variance Threshold \\
\bottomrule
\end{tabular}
\end{table*}

\subsection{Algorithm Pseudocode}
See details in Algorithm~\ref{alg:advisor_trigger}, Algorithm~\ref{alg:pipeline_evaluation} and Algorithm~\ref{alg:main_llapipe_loop}.
\begin{algorithm}
\caption{Adaptive Advisor Triggering Policy}
\label{alg:advisor_trigger}
\begin{algorithmic}[1]
\REQUIRE Episode count $e$, accuracy buffer $\mathcal{B}$, last LLM call $e_{last}$
\ENSURE Triggering decision
\STATE \textbf{// Cooldown period check}
\IF{$e - e_{last} < 5$}
    \RETURN False \COMMENT{Enforce minimum 5-episode cooldown}
\ENDIF
\STATE \textbf{// Buffer readiness check}
\IF{$|\mathcal{B}| < 5$}
    \RETURN False \COMMENT{Insufficient data for trend analysis}
\ENDIF
\STATE \textbf{// Performance trend analysis}
\STATE $\beta \leftarrow$ LinearRegression($\mathcal{B}$)
\IF{$\beta < \theta_{slope} = 0.01$}
    \RETURN True \COMMENT{Performance improvement stagnating}
\ELSE
    \RETURN False \COMMENT{Agent still improving autonomously}
\ENDIF
\end{algorithmic}
\end{algorithm}

\begin{algorithm}
\caption{LLM Pipeline Evaluation and Integration}
\label{alg:pipeline_evaluation}
\begin{algorithmic}[1]
\REQUIRE LLM pipelines $\mathcal{P}_{LLM}$, baseline accuracy $acc_{base}$
\FOR{$P_i \in \mathcal{P}_{LLM}$}
\STATE $acc_i \leftarrow$ Evaluate($P_i$, $D_{val}$)
\IF{$acc_i > acc_{base}$}
\STATE \textbf{// Update Q-table with LLM-discovered knowledge}
\STATE $r_i \leftarrow$ ComputeReward($acc_i$)
\FOR{$(s_t, a_t) \in$ Trajectory($P_i$)}
\STATE $Q(s_t, a_t) \leftarrow Q(s_t, a_t) + \alpha[r + \gamma \max \limits_a Q(s_{t+1}, a)-Q(s_t,a_t)]$
\ENDFOR
\STATE \textbf{break} \COMMENT{Use first improving pipeline}
\ENDIF
\ENDFOR
\end{algorithmic}
\end{algorithm}

\begin{algorithm}
\caption{The LLaPipe Framework}
\label{alg:main_llapipe_loop}
\begin{algorithmic}[1]
\REQUIRE A set of training datasets $\mathcal{D}_{\text{train}}$, operator set $\mathcal{O}$, max episodes $E_{\text{max}}$, max pipeline length $L_{\text{max}}$
\ENSURE An optimized RL agent (Q-network) and a populated Experience Pool $\mathcal{E}$

\STATE Initialize RL Agent (e.g., Q-network) with random weights $\theta$
\STATE Initialize Experience Pool $\mathcal{E}$ (vector database)
\STATE Initialize accuracy buffer $B \leftarrow \emptyset$, last LLM call episode $e_{\text{last}} \leftarrow 0$

\FOR{episode $e = 1$ to $E_{\text{max}}$}
    \STATE Sample a raw dataset $D_0$ from $\mathcal{D}_{\text{train}}$
    \STATE Initialize current pipeline $P \leftarrow \emptyset$, current dataset $D_t \leftarrow D_0$
    
    \FOR{$t = 0$ to $L_{\text{max}}-1$}
        \STATE $s_t \leftarrow \text{ConstructStateVector}(D_t, P)$ \COMMENT{State representation}
        
        \STATE $\text{trigger\_advisor} \leftarrow \text{AdvisorTriggerPolicy}(e, B, e_{\text{last}})$ \COMMENT{Run Algorithm~\ref{alg:advisor_trigger}}
        
        \IF{$\text{trigger\_advisor}$}
            \STATE $e_{\text{last}} \leftarrow e$
            \STATE $A_{\text{suggested}} \leftarrow \text{LLMPolicyAdvisor}(s_t, \mathcal{E})$ \COMMENT{RAG-based LLM call}
            \STATE $a_t \leftarrow \text{SelectActionFromHybridPolicy}(\pi_{RL}, A_{\text{suggested}})$ \COMMENT{Policy Integration}
        \ELSE
            \STATE $a_t \leftarrow \text{SelectActionFromNativePolicy}(s_t)$ \COMMENT{e.g., $\epsilon$-greedy}
        \ENDIF
        
        \IF{$a_t$ is $a_{\text{term}}$}
            \STATE \textbf{break}
        \ENDIF
        
        \STATE $D_{t+1}, r_t \leftarrow \text{ExecuteOperator}(D_t, a_t)$ \COMMENT{Apply operator, get reward}
        \STATE $P \leftarrow P \oplus a_t$
        \STATE Store transition $(s_t, a_t, r_t, s_{t+1})$ in replay buffer for RL agent training
        \STATE $D_t \leftarrow D_{t+1}$
    \ENDFOR
    
    \STATE $\text{acc}_{\text{final}} \leftarrow \text{EvaluateFinalPipeline}(P, D_{\text{val}})$
    \STATE Update accuracy buffer $B$ with $\text{acc}_{\text{final}}$
    
    \IF{$\text{acc}_{\text{final}}$ is high enough}
        \STATE $\text{experience\_entry} \leftarrow \text{CreateExperienceEntry}(D_0, P, \text{acc}_{\text{final}}, \text{trajectory})$
        \STATE Add $\text{experience\_entry}$ to Experience Pool $\mathcal{E}$
    \ENDIF
    
    

\ENDFOR
\end{algorithmic}
\end{algorithm}

\subsection{Full Experimental Results}
\label{app:full_results}

\subsubsection{Extended Comparison Baselines}
\label{app:full_comparison}

To situate LLaPipe within the broadest possible competitive context, we present an extended comparison by integrating our results into the main evaluation table from CtxPipe. Table~\ref{tab:extended_comparison} includes performance data for differentiable methods (DEF, RS, DP-Fix, DP-Flex), SAGA, and other baselines alongside our LLaPipe variants. This direct comparison underscores the competitive performance of our approach.

\begin{table*}
\centering
\caption{Extended comparison of test accuracy, including baselines from CtxPipe. Our methods (Advisor, Advisor\textsuperscript{+}) are added for direct comparison.}
\label{tab:extended_comparison}
\resizebox{\textwidth}{!}{%
\begin{tabular}{lcccccccccc}
\toprule
\textbf{Dataset} & \textbf{DEF} & \textbf{RS} & \textbf{DP-Fix} & \textbf{DP-Flex} & \textbf{DL} & \textbf{HAI-AI} & \textbf{SAGA} & \textbf{CtxPipe} & \textbf{Advisor} & \textbf{Advisor\textsuperscript{+}} \\
\midrule
abalone         & 0.240 & 0.243         & 0.238         & 0.271         & 0.157 & 0.260         & 0.255         & \textbf{0.287}& 0.270         & 0.280 \\
ada\_prior      & 0.848 & 0.844         & \textbf{0.853}& 0.846         & 0.803 & 0.801         & 0.833         & 0.818         & 0.838         & 0.836 \\
avila           & 0.553 & 0.598         & 0.652         & 0.633         & 0.593 & 0.630         & 0.636         & 0.759         & \textbf{0.929}& 0.751 \\
connect-4       & 0.659 & 0.671         & 0.726         & 0.702         & 0.683 & \textbf{0.775}& 0.758         & 0.763         & \textbf{0.775}& \textbf{0.775} \\
eeg             & 0.589 & 0.658         & 0.675         & 0.683         & 0.607 & 0.556         & 0.658         & 0.740         & \textbf{0.839}& 0.811 \\
google          & 0.586 & 0.627         & 0.631         & 0.661         & 0.553 & 0.550         & 0.596         & 0.590         & \textbf{0.675}& 0.674 \\
house           & 0.928 & 0.938         & 0.932         & \textbf{0.952}& 0.771 & 0.928         & 0.913         & 0.818         & 0.908         & 0.908 \\
jungle\_chess   & 0.668 & 0.669         & 0.680         & 0.687         & 0.717 & 0.760         & 0.745         & \textbf{0.861}& \textbf{0.861}& \textbf{0.861} \\
micro           & 0.564 & 0.579         & 0.595         & 0.593         & 0.613 & 0.633         & 0.556         & 0.605         & \textbf{0.643}& 0.633 \\
mozilla4        & 0.855 & 0.922         & 0.924         & 0.927         & 0.747 & 0.870         & 0.932         & 0.940         & 0.937         & \textbf{0.944} \\
obesity         & 0.775 & 0.841         & \textbf{0.891}& 0.874         & 0.590 & 0.768         & 0.751         & 0.868         & 0.865         & 0.835 \\
page-blocks     & 0.942 & 0.959         & 0.959         & \textbf{0.973}& 0.940 & 0.935         & 0.849         & 0.965         & 0.963         & 0.961 \\
pbcseq          & 0.710 & 0.730         & 0.728         & 0.725         & 0.680 & 0.733         & \textbf{0.866}& 0.805         & 0.753         & 0.753 \\
pol             & 0.884 & 0.879         & 0.903         & 0.916         & 0.873 & 0.916         & 0.888         & 0.949         & \textbf{0.981}& \textbf{0.981} \\
run\_or\_walk   & 0.719 & 0.829         & 0.903         & 0.912         & 0.820 & 0.915         & 0.832         & 0.956         & 0.953         & \textbf{0.976} \\
shuttle         & 0.964 & 0.996         & 0.998         & 0.999         & 0.790 & 0.951         & 0.405         & \textbf{1.000}& 0.987         & 0.986 \\
uscensus        & 0.848 & 0.840         & \textbf{0.854}& 0.852         & 0.813 & 0.807         & 0.835         & 0.845         & 0.846         & 0.847 \\
wall-robot-nav  & 0.697 & 0.872         & 0.905         & 0.913         & 0.927 & 0.896         & 0.841         & 0.946         & \textbf{0.961}& 0.946 \\
\midrule
\textbf{AVERAGE}& 0.724 & 0.761         & 0.780         & 0.784         & 0.704 & 0.760         & 0.731         & 0.806         & \textbf{0.833}& \textbf{0.820} \\
\textbf{RANK}   & 7.81  & 6.56          & 4.86          & 4.19          & 8.44  & 6.58          & 6.75          & 3.61          & \textbf{3.00} & \textbf{3.19} \\
\bottomrule
\end{tabular}
}
\end{table*}


\begin{table}
    \centering
    \caption{Comparison of test accuracy between DQN and DQN with Advisor\textsuperscript{+} (Adv\textsuperscript{+})}
\begin{tabular}{lcc|lcc}
\toprule
\textbf{Dataset} & \textbf{DQN} & \textbf{Adv\textsuperscript{+}} & \textbf{Dataset} & \textbf{DQN} & \textbf{Adv\textsuperscript{+}} \\
\midrule
abalone & 0.257 & \textbf{0.264} & page & 0.952 & \textbf{0.963} \\
ada\_prior & 0.812 & 0.812 & pbcseq & 0.735 & 0.735 \\
avila & 0.620 & 0.620 & pol & 0.796 & \textbf{0.969} \\
connect-4 & 0.664 & \textbf{0.747} & run & 0.972 & 0.863 \\
eeg & 0.613 & 0.613 & shuttle & 0.965 & \textbf{0.978} \\
google & 0.594 & 0.594 & uscensus & 0.832 & \textbf{0.845} \\
house & 0.908 & 0.908 & wall & 0.853 & \textbf{0.875} \\
jungles & 0.697 & 0.695 & micro & 0.589 & \textbf{0.625} \\
mozilla4 & 0.871 & 0.871 & obesity & 0.825 & 0.825 \\
\midrule
\multicolumn{1}{l}{\textbf{AVERAGE}} & 
\multicolumn{1}{c}{0.753} & 
\multicolumn{1}{c}{\textbf{0.767}} & 
\multicolumn{1}{l}{\textbf{RANK}} & 
\multicolumn{1}{c}{6.28} &
\multicolumn{1}{c}{\textbf{5.78}} \\
\bottomrule
\end{tabular}
    \label{tab:dqn_advisor_compare}
\end{table}


\begin{table}
    \centering
    \caption{Comparison between CtxPipe (Ctx) and CtxPipe with Advisor\textsuperscript{+} (Adv\textsuperscript{+})}
    \begin{tabular}{lcc|lcc}
\toprule
\textbf{Dataset} & \textbf{Ctx} & \textbf{Adv\textsuperscript{+}} & \textbf{Dataset} & \textbf{Ctx} & \textbf{Adv\textsuperscript{+}} \\
\midrule
eeg & 0.740 & 0.743 & google & 0.590 & \textbf{0.616} \\
micro & 0.605 & 0.607 & mozilla4 & 0.940 & \textbf{0.944} \\
page & 0.965 & 0.966 & run & 0.956 & \textbf{0.967} \\
\bottomrule
    \end{tabular}
    \label{tab:ctx_advisor_compare}
\end{table}

\subsubsection{Detailed Pipeline Composition}
\label{app:pipeline_details}

To provide deeper insight into the qualitative differences between the pipelines discovered by CtxPipe and our LLaPipe variants, we present the detailed composition of the best-performing pipeline for each method on every dataset. Table~\ref{tab:operator_mapping} provides the mapping from the operator IDs used in Table~\ref{tab:pipeline_composition} to their corresponding names.

\begin{table}
\centering
\caption{Mapping of operator IDs to their names, based on our operator set.}
\label{tab:operator_mapping}
\begin{tabular}{ll|ll}
\toprule
\textbf{ID} & \textbf{Operator Name} & \textbf{ID} & \textbf{Operator Name} \\
\midrule
0 & ImputerCatPrim & 13 & Normalizer \\
1 & ImputerMean & 14 & KBinsDiscretizerOrdinal \\
2 & ImputerMedian & 15 & PolynomialFeatures \\
3 & ImputerNum & 16 & InteractionFeatures \\
4 & LabelEncoder & 17 & PCA\_AUTO \\
5 & OneHotEncoder & 18 & PCA\_LAPACK \\
6 & MinMaxScaler & 19 & PCA\_ARPACK \\
7 & MaxAbsScaler & 20 & IncrementalPCA \\
8 & RobustScaler & 21 & KernelPCA \\
9 & StandardScaler & 22 & TruncatedSVD \\
10 & QuantileTransformer & 23 & RandomTreesEmbedding \\
11 & LogTransformer & 24 & VarianceThreshold \\
12 & PowerTransformer & -1 & BlankOperation \\
\bottomrule
\end{tabular}
\end{table}

Table~\ref{tab:pipeline_composition} details the generated pipelines and their resulting accuracy scores. The `Pipe Length` row at the bottom summarizes the average number of operators in the pipelines generated by each method. This table illustrates that LLaPipe not only achieves competitive or superior accuracy but often does so with different, and sometimes more complex or novel, pipeline structures.

\begin{table*}
\centering
\caption{Detailed composition and accuracy of pipelines generated by CtxPipe and LLaPipe variants. Numbers in brackets correspond to operator IDs in Table~\ref{tab:operator_mapping}.}
\label{tab:pipeline_composition}
\resizebox{\textwidth}{!}{%
\begin{tabular}{l|ll|ll|ll}
\toprule
\textbf{Dataset} & \multicolumn{2}{c|}{\textbf{CtxPipe}} & \multicolumn{2}{c|}{\textbf{Advisor}} & \multicolumn{2}{c}{\textbf{Advisor\textsuperscript{+}}} \\
& \textbf{Accuracy} & \textbf{Pipeline} & \textbf{Accuracy} & \textbf{Pipeline} & \textbf{Accuracy} & \textbf{Pipeline} \\
\midrule
abalone       & 0.287 & [-1, -1, 4, 24, 12, 15]     & 0.270 & [6, 9, 11, 16, 20]   & 0.273 & [8, 15]		\\
ada\_prior    & 0.818 & [-1, -1, 4, -1, 6, 23]      & 0.838 & [5, 12, 9]           & 0.841 & [12, 6, 5]		\\
avila         & 0.759 & [-1, -1, -1, 24, 12, 23]    & 0.929 & [10, 23, 18, 15]     & 0.751 & [23]		   \\
connect-4     & 0.763 & [-1, -1, -1, -1, 12, 23]    & 0.775 & [15]			       & 0.775 & [15, 4]			   \\
eeg           & 0.740 & [-1, -1, -1, 23, 24, 12]    & 0.839 & [18, 3, 15, 9]	   & 0.811 & [10, 15]		   \\
google        & 0.590 & [1, -1, 4, 6, 23, 24]       & 0.675 & [2, 10, 9, 15]	   & 0.674 & [2, 11, 8, 15]	   \\
house         & 0.818 & [1, -1, 4, 12, 23, 24]      & 0.908 & [1]				   & 0.908 & [1]				   \\
jungle\_chess & 0.861 & [-1, -1, -1, 24, 6, 23]     & 0.861 & [23]			       & 0.861 & [23]			   \\
micro         & 0.605 & [-1, -1, -1, -1, 6, 23]     & 0.643 & [12, 15, 18]	       & 0.633 & [15]			   \\
mozilla4      & 0.940 & [-1, -1, -1, 6, 23, 24]     & 0.937 & [13, 10, 24, 15]     & 0.944 & [12, 6, 23]		   \\
obesity       & 0.868 & [-1, -1, 4, 24, 6, 23]      & 0.865 & [10, 5, 9, 15]	   & 0.835 & [0, 15, 17]		   \\
page-blocks   & 0.965 & [-1, -1, -1, -1, 6, 23]     & 0.965 & [3, 10, 15]		   & 0.961 & [11, 20]		   \\
pbcseq        & 0.805 & [1, -1, -1, -1, 6, 23]      & 0.753 & [23, 6]			   & 0.753 & [23]		   \\
pol           & 0.949 & [-1, -1, -1, 8, 23, 24]     & 0.981 & [15]                 & 0.981 & [16]		   \\
run\_or\_walk & 0.956 & [-1, -1, -1, 24, 12, 15]    & 0.953 & [12, 15, 17]	       & 0.976 & [14, 23]		   \\
shuttle       & 1.000 & [-1, -1, -1, 23, 24, 12]    & 0.987 & [9, 15]			   & 0.986 & [10, 5]		   \\
uscensus      & 0.845 & [-1, -1, 4, -1, 6, 23]      & 0.846 & [2, 5, 7]		       & 0.847 & [5, 9]			   \\
wall-robot-nav& 0.946 & [-1, -1, -1, -1, 6, 23]     & 0.961 & [10, 9, 15]		   & 0.946 & [23]		   \\
\midrule
\textbf{Pipe Length} & \multicolumn{2}{c|}{\textbf{6}} & \multicolumn{2}{c|}{\textbf{2.83}} & \multicolumn{2}{c}{\textbf{1.89}} \\
\bottomrule
\end{tabular}
}
\end{table*}

\subsubsection{Setup Ranking}
\label{app:setup_rank}
\begin{figure}
    \centering
    \includegraphics[width=0.5\linewidth]{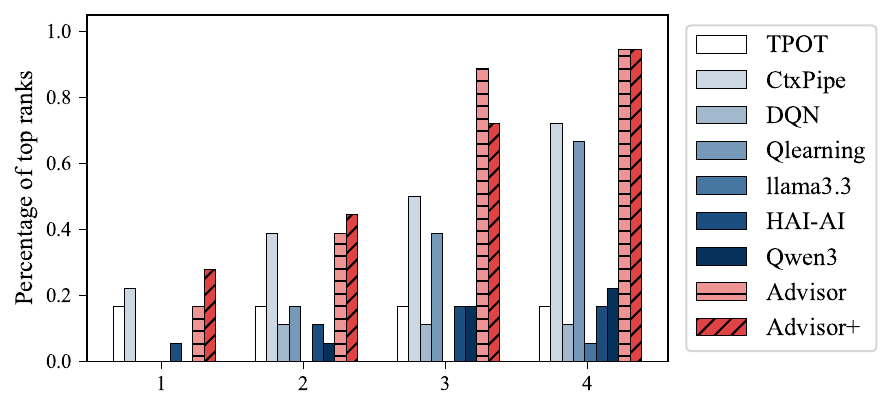}
    \caption{Top-k Accuracy Ranks}
    \label{fig:ranks}
\end{figure}

Since the efficacy of data preparation varies across the datasets, we further conduct a granular analysis by ranking the competitors’ effectiveness at the dataset level. Across 18 diverse datasets, LLaPipe consistently finds higher quality pipelines, achieving the best average rank among all tested methods, as Figure~\ref{fig:ranks}. 

\subsection{LLM Prompt Design}
\label{app:prompt_design}
\subsubsection{A Concrete Prompt Example}
Here is a concrete example of a filled-out prompt for the \texttt{avila} dataset at an early stage of pipeline construction. This demonstrates how the abstract template is instantiated with real data and retrieved experiences.

\begin{tcolorbox}[
    colback=promptbackground,      %
    colframe=promptborder,         %
    coltitle=black,                %
    fonttitle=\bfseries,           %
    title=LLM Prompt Example: \texttt{avila} Dataset, %
    arc=2mm,                       %
    boxrule=1pt,                   %
    left=6mm,
    right=6mm,
    top=3mm,
    bottom=3mm,
    listing only,                  %
    listing options={
        language=bash,             %
        basicstyle=\ttfamily\small,%
        keywordstyle=\color{blue},
        commentstyle=\color{gray},
        stringstyle=\color{red!60!black}
    }
]

You are an expert data scientist specializing in automated machine learning. Your task is to analyze the provided dataset context and historical information to propose a list of 1 to 3 optimal data preparation pipelines. Each pipeline should be a sequence of data processing operators that aims to maximize the prediction accuracy of a downstream classification model.

For each proposed pipeline, you must provide: 

(1) A list of operator names in sequence. (2) A confidence score (from 0.0 to 1.0) for your suggestion. (3) A brief, clear rationale explaining why this pipeline is suitable for the given data.

\textbf{Current situation and context}:

- Task Type: Logistic regression classification

- Key Dataset Statistics:
  
\quad- Size of dataset: 16693 rows, 10 cols
  
\quad- Missing Values: No
  
\quad- Feature Types: All numerical

\quad- Cols with skewed distribution: f1, f3, f4

\quad- Cols with outliers: f0 (9.50\%), f2 (10.50\%), f7 (2.62\%)

- Current Partial Pipeline: [PowerTransformer, RobustScaler, ...]

\textbf{Available operators}: ImputerMean, StandardScaler, QuantileTransformer, PowerTransformer, PCA, 
MinMaxScaler, VarianceThreshold, ...

---

\textbf{[Example 1]}

-  \textit{Context}: A dataset with no missing values but high skewed distribution columns (B3, B9, ...), oulier columns (B1, B2, B5, ...)

-  \textit{Pipeline}: [QuantileTransformer, RandomTrees-Embedding, MinMaxScaler], accuracy 0.92


\textbf{[Example 2]}

- \textit{Context}: A dataset with many correlated numerical features (A5 and A8: correlation 92\%, A5 and A9: correlation 88\%, ...)

-  \textit{Pipeline}: [StandardScaler, PCA], accuracy 0.88


\end{tcolorbox}

\subsection{Hyperparameter Settings}
\label{app:hyperparameters}

The reproducibility of our experiments relies on a precise configuration of hyperparameters. Table~\ref{tab:hyperparameters} lists the key settings for our LLaPipe framework, which is built upon a traditional, tabular Q-Learning agent. A crucial aspect of our implementation is the discretization of the continuous state space into a finite set of states to make it compatible with tabular Q-Learning. The settings were kept consistent across all datasets.

\begin{table*}[h!]
\centering
\caption{Hyperparameter settings for the LLaPipe framework with a Tabular Q-Learning agent.}
\label{tab:hyperparameters}
\begin{tabular}{llc}
\toprule
\textbf{Module} & \textbf{Hyperparameter} & \textbf{Value} \\
\midrule
\multicolumn{3}{c}{\textbf{RL Agent (Tabular Q-Learning)}} \\
\midrule
\textit{Learning} & Learning Rate ($\alpha$) & 1 \\
& Discount Factor ($\gamma$) & 0.9 \\
\addlinespace
\textit{Exploration} & $\epsilon$-Greedy Decay Strategy & Exponential \\
& Initial Epsilon ($\epsilon_{\text{start}}$) & 1.0 \\
& Final Epsilon ($\epsilon_{\text{end}}$) & 0.1 \\
& Epsilon Decay Factor & 0.99 \\
\midrule
\multicolumn{3}{c}{\textbf{LLM Policy Advisor}} \\
\midrule
\textit{Model} & LLM Model & LLaMA-3.3-70B \\
& Temperature & 0.1 \\
\addlinespace
\textit{Retrieval} & Number of Retrieved Examples ($k$) & 3 \\
& Text Embedding Model & nomic-embed-text \\
& Vector Database Index & FAISS (IndexFlatL2) \\
\midrule
\multicolumn{3}{c}{\textbf{Adaptive Advisor Triggering (Advisor\textsuperscript{+})}} \\
\midrule
\textit{Triggering} & Slope Threshold ($\theta_{slope}$) & 0.01 \\
& Performance Buffer Size ($|\mathcal{B}|$) & 10 episodes \\
& Cooldown Period & 5 episodes \\
\midrule
\multicolumn{3}{c}{\textbf{Framework Settings}} \\
\midrule
\textit{General} & Max Pipeline Length ($L_{\text{max}}$) & 9 \\
& Total Training Episodes ($E_{\text{max}}$) & 100 \\
\addlinespace
\textit{Random Seed} & Dataset Train Test Split & 0 \\
\bottomrule
\end{tabular}
\end{table*}

\subsection{Theorem}

\subsection{Proof of Theorem \ref{thm:linear_approx} (Local Linear Approximation)}
\begin{proof}
Let $V^\pi(s)$ be the value function (expected cumulative reward) for a policy $\pi$ starting from state $s$. The expected accuracy after an episode is monotonically related to this value. Consider a small update to the policy, from $\pi$ to $\pi'$, such that $\pi' = \pi + \Delta\pi$. A first-order Taylor expansion of the value function gives:
\begin{equation}
    V^{\pi'}(s) \approx V^\pi(s) + \nabla_\pi V^\pi(s) \cdot \Delta\pi
\end{equation}
Assuming that each episode's learning step corresponds to a small, incremental policy update $\Delta\pi$ in a consistent direction, the cumulative change in value after $e$ episodes is approximately linear. Since accuracy is a function of the final value, we can write $\mathbb{E}[\text{acc}(e)] \approx f(V^{\pi_e}(s))$, which simplifies to a linear form $\text{acc}_0 + \beta \cdot e$ for a short series of episodes where the gradient $\nabla_\pi V^\pi(s)$ is relatively constant. This justifies using a linear regression model as a local approximation of the learning trend.
\end{proof}

\subsection{Proof of Theorem \ref{thm:slope_optimality} (Optimality of Slope-Based Triggering)}
\begin{proof}
We frame the decision as an optimal stopping problem. At any episode, the agent can choose one of two actions: (1) \textbf{Continue} with RL, or (2) \textbf{Trigger} the LLM. Let $C_{\text{RL}}$ be the cost of one RL episode and $C_{\text{LLM}}$ be the cost of an LLM invocation and subsequent evaluation. Let $\mathbb{E}[\Delta\text{acc}_\text{{LLM}}]$ be the expected immediate accuracy gain from a successful LLM intervention.

The value of continuing for a small number of future episodes $\Delta e$ under the current learning slope $\beta$ is:
\begin{equation}
    V_{\text{continue}}(\beta, \Delta e) = \sum_{i=1}^{\Delta e} (\beta \cdot i - C_\text{{RL}})
\end{equation}
The value of triggering the LLM now is:
\begin{equation}
    V_{\text{LLM}} = \mathbb{E}[\Delta\text{acc}_{LLM}] - C_{LLM}
\end{equation}
The optimal policy is to trigger the LLM if $V_{\text{LLM}} > V_{\text{continue}}$. For a one-step horizon ($\Delta e = 1$), this simplifies to triggering if $\mathbb{E}[\Delta\text{acc}_{LLM}] - C_{LLM} > \beta - C_{RL}$.
This defines a threshold $\theta_{slope} = \mathbb{E}[\Delta\text{acc}_{LLM}] - C_{LLM} + C_{RL}$. For typical cost structures where $C_{LLM} \gg C_{RL}$ and the expected gain is a small positive constant (e.g., 0.05), this leads to a small positive slope threshold (e.g., $\theta_{slope} \approx 0.01$). Triggering when $\beta < \theta_{slope}$ is therefore the optimal strategy to maximize the cost-adjusted performance gain.
\end{proof}

\subsection{Proof of Theorem \ref{thm:first_improvement} (Optimality of First-Improvement Sampling)}
\begin{proof}
Let the LLM generate $k$ candidate pipelines $\{P_1, \dots, P_k\}$, ordered by the LLM's confidence, such that the expected quality $\mathbb{E}[\text{acc}_i]$ is non-increasing with $i$. Let $p_i = \text{Pr}(\text{acc}_i > \text{acc}_{\text{base}})$ be the probability that pipeline $P_i$ is an improvement, and let $C_{\text{eval}}$ be the constant cost of evaluating one pipeline.

The "first-improvement" strategy stops at the first $i$ where $\text{acc}_i > \text{acc}_{\text{base}}$. The expected number of evaluations is $\mathbb{E}[N] = \sum_{i=1}^k i \cdot p_i \cdot \prod_{j=1}^{i-1}(1-p_j)$. The expected total improvement is $\mathbb{E}[\Delta\text{acc}] = \sum_{i=1}^k \mathbb{E}[\text{acc}_i - \text{acc}_{\text{base}} | \text{acc}_i > \text{acc}_{\text{base}}] \cdot p_i \cdot \prod_{j=1}^{i-1}(1-p_j)$.

The objective is to maximize the value-per-cost ratio, $\rho = \frac{\mathbb{E}[\Delta\text{acc}]}{\mathbb{E}[N] \cdot C_{\text{eval}}}$. Due to the ordering of pipelines by expected quality, the potential gain from evaluating further pipelines (i.e., $P_{i+1}, \dots, P_k$) given that the first $i$ have failed to improve, is progressively lower. The first-improvement stopping rule is a known optimal strategy for this class of problems (related to the "secretary problem"), as it terminates as soon as a positive gain is realized, avoiding further evaluation costs on potentially lower-quality candidates.
\end{proof}

\subsection{Proof of Theorem \ref{thm:cost_reduction} (Expected Cost Reduction)}
\begin{proof}
Let $T$ be the total number of training episodes. Let $C_{LLM}$ be the cost of an LLM call and $C_{RL}$ be the cost of a standard RL episode.

The total cost of a \textbf{fixed-frequency strategy} that calls the LLM every $K$ episodes is:
\begin{equation}
    \text{Cost}_{\text{fixed}} = \frac{T}{K} \cdot C_{LLM} + \left(T - \frac{T}{K}\right) \cdot C_{RL}
\end{equation}
Let $p_{stag}$ be the probability that the agent's learning is stagnated (i.e., $\beta < \theta_{slope}$). This is the probability that Advisor\textsuperscript{+} will trigger the LLM.

The total expected cost of the \textbf{Advisor\textsuperscript{+} strategy} is:
\begin{equation}
    \mathbb{E}[\text{Cost}_{\text{adaptive}}] = (T \cdot p_{stag}) \cdot C_{LLM} + (T \cdot (1-p_{stag})) \cdot C_{RL}
\end{equation}
The expected cost reduction, $\Delta\text{Cost} = \text{Cost}_{\text{fixed}} - \mathbb{E}[\text{Cost}_{\text{adaptive}}]$, can be substantial. For simplicity, let's compare to a naive strategy of calling the LLM every episode ($K=1$).
\begin{align}
    \Delta\text{Cost} 
    &= T \cdot (1 - p_{stag}) \cdot C_{LLM} - T \cdot (1 - p_{stag}) \cdot C_{RL} \\
    &= T \cdot (1 - p_{stag}) \cdot (C_{LLM} - C_{RL})
\end{align}
Since $C_{LLM} \gg C_{RL}$ and typically $p_{stag} < 1$, the cost reduction is significant and positive. The same logic applies for any fixed $K>1$.
\end{proof}

\end{document}